\documentstyle[12pt,aaspp4]{article}   
\lefthead{Edelson et al.} 
\righthead{Multiwavelength Variability in NGC 4151}
\newcommand{\rf}{\reference}

\newcommand{\gs}{\mbox{$\stackrel {>}{_{\sim}}$}}
\newcommand{\ls}{\mbox{$\stackrel {<}{_{\sim}}$}}
\newcommand{\et}{et al.\ }
\newcommand{\IUE}{{\it IUE}}
\newcommand{\ROSAT}{{\it ROSAT}}
\newcommand{\ASCA}{{\it ASCA}}
\newcommand{\CGRO}{{\it CGRO}}
\def\n{\footnotemark}
\begin{document}
\normalsize
\begin{center}
{\bf MULTIWAVELENGTH OBSERVATIONS OF }\\
{\bf SHORT TIME-SCALE VARIABILITY IN NGC~4151.}\\
{\bf IV.\ ANALYSIS OF MULTIWAVELENGTH CONTINUUM VARIABILITY}\\
\end{center}
\vskip 0.4cm
\small
\begin{center}
\sc
R.A.~Edelson,\n\
T.~Alexander,\n\
D.M.~Crenshaw,\n\
S.~Kaspi,\footnotemark[2]\
M.A.~Malkan,\n\
B.M.~Peterson,\n\
R.S.~Warwick,\n\
J.~Clavel,\n\
A.V.~Filippenko,\n\
K.~Horne,\n\
K.T.~Korista,\n\
G.A.~Kriss,\n\
J.H.~Krolik,\footnotemark[11]\
D.~Maoz,\footnotemark[2]\
K.~Nandra,\n\
P.T.~O'Brien,\footnotemark[6]\
S.V.~Penton,\n\
T.~Yaqoob,\footnotemark[12]\
P.~Albrecht,\n\
D.~Alloin,\n\
T.R.~Ayres,\footnotemark[13]\
T.J.~Balonek,\n\
P.~Barr,\footnotemark[7]\
A.J.~Barth,\footnotemark[8]\
R.~Bertram,\footnotemark[5]$^{,}$\n\
G.E.~Bromage,\footnotemark[6]\
M.~Carini,\n\
T.E.~Carone,\n$^{,}$\n\
F.-Z.~Cheng,\n\
K.K.~Chuvaev,\n$^{,}$\n\
M.~Dietrich,\n\
D.~Dultzin-Hacyan,\n\
C.M.~Gaskell,\n\
I.S.~Glass,\n\
M.R.~Goad,\n\
S.~Hemar,\footnotemark[2]\
L.C.~Ho,\n\
J.P.~Huchra,\footnotemark[29]\
J.~Hutchings,\n\
W.N.~Johnson,\n\
D.~Kazanas,\footnotemark[12]\
W.~Kollatschny,\footnotemark[14]\
A.P.~Koratkar,\footnotemark[28]\
O.~Kovo,\footnotemark[2]\
A.~Laor,\n\
G.M.~MacAlpine,\n\
P.~Magdziarz,\n\
P.G.~Martin,\n\
T.~Matheson,\footnotemark[8]\
B.~McCollum,\footnotemark[18]\
H.R.~Miller,\n\
S.L.~Morris,\footnotemark[30]\
V.L.~Oknyanskij,\n\
J.~Penfold,\n\
E.~P\'{e}rez,\footnotemark[28]\
G.C.~Perola,\n\
G.~Pike,\footnotemark[12]$^{,}$\n\
R.W.~Pogge,\footnotemark[5]\
R.L.~Ptak,\n\
B.-C.~Qian,\n\
M.C.~Recondo-Gonz\'{a}lez,\n\
G.A.~Reichert,\n\
J.M.~Rodr\'{\i}guez-Espinoza,\n\
P.M.~Rodr\'{\i}guez-Pascual,\n\
E.L.~Rokaki,\n\
J.~Roland,\n\
A.C.~Sadun,\n\
I.~Salamanca,\n\
M.~Santos-Lle\'{o},\n\
J.C.~Shields,\n$^{,}$\n\
J.M.~Shull,\footnotemark[13]$^{,}$\n\
D.A.~Smith,\footnotemark[6]\
S.M.~Smith,\footnotemark[5]\
M.A.J.~Snijders,\n$^{,}$\n\
G.M.~Stirpe,\n\
R.E.~Stoner,\footnotemark[41]\
W.-H.~Sun,\n\
M.-H.~Ulrich,\n\
E.~van~Groningen,\n\
R.M.~Wagner,\footnotemark[5]$^{,}$\footnotemark[17]\
S.~Wagner,\footnotemark[24]\
I.~Wanders,\footnotemark[5]\
W.F.~Welsh,\n\
R.J.~Weymann,\n\
B.J.~Wilkes,\footnotemark[29]\
H.~Wu,\n\
J.~Wurster,\footnotemark[1]\
S.-J.~Xue,\footnotemark[21]\
A.A.~Zdziarski,\n\
W.~Zheng,\footnotemark[11]\
and Z.-L.~Zou\footnotemark[63]\
\end{center}
\vskip 0.4cm
\footnotesize
\leftskip=1.em
\parindent=-1.em

\setcounter{footnote}{0}

\n Department of Physics and Astronomy, 203 Van Allen Hall, University of 
Iowa, Iowa City, IA 52242. Electronic mail: edelson@spacly.physics.uiowa.edu

\n School of Physics and Astronomy and the Wise Observatory,
The Raymond and Beverly Sackler Faculty of Exact Sciences,
Tel-Aviv University, Tel-Aviv 69978, Israel.

\n Astronomy Program, Computer Sciences Corporation,
NASA Goddard Space Flight Center, Code 681,
Greenbelt, MD  20771.

\n Department of Astronomy, University of California, 
Math-Science Building, Los Angeles, CA 90024.

\n Department of Astronomy, The Ohio State University,
174 West 18th Avenue, Columbus, OH 43210.

\n Department of Astronomy, University of Leicester,
University Road, Leicester LE1\,7RH, United Kingdom 

\n ISO Project, European Space Agency, Apartado 50727,
28080 Madrid, Spain.

\n Department of Astronomy, University of California,
Berkeley, CA  94720.

\n School of Physics and Astronomy, University of St.\ Andrews,
North Haugh, St.\ Andrews KY16\,9SS, Scotland, United Kingdom.

\n  Department of Physics and Astronomy,
University of Kentucky, Lexington, KY  40506.

\n Department of Physics and Astronomy, The Johns Hopkins
University, Baltimore, MD 21218.

\n Laboratory for High Energy Astrophysics,
NASA Goddard Space Flight Center, Greenbelt, MD 20771.

\n Center for Astrophysics and Space Astronomy,
University of Colorado, Campus Box 389, Boulder, CO 80309.

\n Universit\"{a}ts-Sternwarte G\"{o}ttingen,
Geismarlandstrasse 11, D-37083 G\"{o}ttingen, Germany.

\n Centre d'Etudes de Saclay, Service d'Astrophysique,
Orme des Merisiers, 91191 Gif-Sur-Yvette Cedex, France.
 
\n Department of Physics and Astronomy, Colgate University,
Hamilton, NY 13346.

\n Mailing address: Lowell Observatory, 1400 West Mars Hill Road,
Flagstaff, AZ  86001.

\n Computer Sciences Corporation, NASA Goddard Space Flight
Center, Code 684.9, Greenbelt, MD 20771.

\n Space Sciences Laboratory, University of California,
Berkeley, CA 94720, and Eureka Scientific, Inc.

\n Current address: 28740 W.\ Fox River Dr., Cary, IL 60013. 

\n Center for Astrophysics, University of Science and Technology,
Hefei, Anhui, People's Republic of China.

\n Crimean Astrophysical Observatory, P/O Nauchny,
334413 Crimea, Ukraine.

\n Deceased, 1994 November 15.

\n Landessternwarte,  K\"{o}nigstuhl,
D-69117 Heidelberg, Germany.

\n Universidad  Nacional Autonoma de Mexico, Instituto de Astronomia,
Apartado Postal 70-264, 04510 Mexico D.F., Mexico.

\n Department of Physics and Astronomy,
University of Nebraska, Lincoln, NE 68588.

\n South African Astronomical Observatory, P.O.\ Box 9,
Observatory 7935, South Africa.

\n Space Telescope Science Institute,
3700 San Martin Drive, Baltimore, MD 21218.

\n Harvard-Smithsonian Center for Astrophysics, 
60 Garden Street, Cambridge, MA  02138.

\n Dominion Astrophysical Observatory, 
5071 West Saanich Road, Victoria, B.C. V8X 4M6, Canada.

\n Naval Research Laboratory, Code 4151, 4555 Overlook SW,
Washington, DC 20375-5320.

\n Physics Department, Technion-Israel Institute of Technology,
Haifa 32000, Israel.

\n Department of Astronomy, University of Michigan, Dennison Building, Ann
Arbor, MI  48109.

\n Astronomical Observatory, Jagiellonian University, Orla 171,
30-244 Cracow, Poland.

\n Canadian Institute for Theoretical Astrophysics,
University of Toronto, Toronto, ON M5S 1A1, Canada.

\n Department of Physics and
Astronomy, Georgia State University, Atlanta, GA  30303.

\n Sternberg Astronomical Institute, University of Moscow,
Universitetskij Prosp.\ 13, Moscow 119899, Russia.

\n Department of Physics and Astronomy, University of Calgary,
2500 University Drive NW, Calgary, AB T2N 1N4, Canada, and
Department of Mathematics, Physics, and Engineering, 
Mount Royal College, Calgary T3E 6K6, Canada.

\n Istituto Astronomico dell'Universit\`{a}, Via Lancisi 29, I-00161 Rome,
Italy.

\n Mailing address: 816 S.\ LaGrange Rd., LaGrange, IL 60525.

\n Department of Physics and Astronomy, Bowling Green State University, 
Bowling Green, OH  43403.

\n Shanghai Observatory, Chinese Academy of Sciences, People's Republic of
China.

\n Facultad de Ciencias, Dept.\ F\'{\i}sicas, Universidad de Oviedo,
C/ Calvo Sotelo, s/n. 
Oviedo, Asturias, Spain.

\n NASA Goddard Space Flight Center, Code 631,
Greenbelt, MD 20771.

\n Instituto de Astrof\'{\i}sica de Canarias, E-38200 La Laguna, Tenerife,
Spain.

\n ESA IUE Observatory, P.O.\ Box 50727,
28080 Madrid, Spain.

\n Royal Observatory Edinburgh, University of Edinburgh, Blackford Hill,
Edinburgh EH9\,3HJ, United Kingdom.

\n Institut d'Astrophysique, 98 bis Boulevard Arago,
F-75014 Paris, France.

\n Department of Physics and Astronomy and Bradley Observatory, Agnes Scott
College, Decatur, GA 30030.

\n Royal Greenwich Observatory, Madingley Road,
Cambridge CB3 0EZ, United Kingdom.

\n LAEFF, Apdo.\ 50727, E-28080 Madrid, Spain.

\n Steward Observatory, University of Arizona, Tucson, AZ  85726.

\n Hubble Fellow.

\n JILA;
University of Colorado and
National Institute of Standards and Technology, 
Campus Box 440, Boulder, CO 80309.

\n IRAM, 300 Rue de la Piscine, 38046 Saint Martin d'Heres,
France.

\n Mailing address: rue Elysee Reclubus 1 bis., 38100 Grenoble, France.

\n Osservatorio Astronomico di Bologna, Via Zamboni 33, I-40126,
Bologna, Italy.

\n Institute of Astronomy, National Central University, 
Chung-Li, Taiwan 32054, Republic of China.

\n European Southern Observatory, Karl Schwarzschild Strasse 2,
85748 Garching, Germany.

\n Astronomiska observatoriet, Box 515, S-751 20 Uppsala, Sweden.

\n Department of Physics, Keele University, Keele ST5 5BG,
Staffordshire, United Kingdom.

\n Observatories of the Carnegie Institution of Washington,
813 Santa Barbara Street, Pasadena, CA  91101.

\n Beijing Astronomical Observatory, Chinese Academy of Sciences,
Beijing 100080, People's Republic of China.

\n N.\ Copernicus Astronomical Center, Bartycka 18, 00-716 Warsaw, Poland.

\leftskip=0em
\parindent=1.5em

\normalsize

\begin{abstract} 

This paper combines data from the three preceding papers in order to analyze 
the multi-waveband variability and spectral energy distribution of the 
Seyfert~1 galaxy NGC~4151 during the December 1993 monitoring campaign.
The source, which was near its peak historical brightness, showed strong, 
correlated variability at X-ray, ultraviolet, and optical wavelengths.
The strongest variations were seen in medium energy ($\sim$1.5~keV) X-rays, 
with a normalized variability amplitude (NVA) of 24\%.
Weaker (NVA = 6\%) variations (uncorrelated with those at lower energies) 
were seen at soft $\gamma$-ray energies of $\sim$100~keV.
No significant variability was seen in softer (0.1--1~keV) X-ray bands.
In the ultraviolet/optical regime, the NVA decreased from 9\% to 1\% as the 
wavelength increased from 1275~\AA\ to 6900~\AA.
These data do not probe extreme ultraviolet (1200~\AA\ to 0.1~keV) or hard 
X-ray (2--50~keV) variability.
The phase differences between variations in different bands were consistent 
with zero lag, with upper limits of $\ls$0.15~day between 1275~\AA\ and the 
other ultraviolet bands, $\ls$0.3~day between 1275~\AA\ and 1.5~keV, and 
$\ls$1~day between 1275~\AA\ and 5125~\AA. 
These tight limits represent more than an order of magnitude improvement 
over those determined in previous multi-waveband AGN monitoring campaigns.
The ultraviolet fluctuation power spectra showed no evidence for periodicity, 
but were instead well-fitted with a very steep, red power-law ($ a = -2.5 $).

If photons emitted at a ``primary" waveband are absorbed by nearby material 
and ``reprocessed" to produce emission at a secondary waveband, causality 
arguments require that variations in the secondary band follow those in the 
primary band.
The tight interband correlation and limits on the ultraviolet and medium 
energy X-ray lags indicate that the reprocessing region is smaller than 
$\sim$0.15~lt-day in size.
After correcting for strong (factor of $\gs$15) line of sight absorption, the 
medium energy X-ray luminosity variations appear adequate to drive the 
ultraviolet/optical variations.
However, the medium energy X-ray NVA is 2--4 times that in the ultraviolet, 
and the single-epoch, absorption-corrected X-ray/$\gamma$-ray luminosity is
only about 1/3 that of the ultraviolet/optical/infrared, suggesting that at
most $\sim$1/3 of the total low-energy flux could be reprocessed high-energy 
emission. 

The strong wavelength dependence of the ultraviolet NVAs is consistent with 
an origin in an accretion disk, with the variable emission coming from the 
hotter inner regions and non-variable emission from the cooler outer regions. 
These data, when combined with the results of disk fits, indicate a boundary 
between these regions near a radius of order $ R \approx 0.07 $~lt-day.
No interband lag would be expected as reprocessing (and thus propagation 
between regions) need not occur, and the orbital time scale of $\sim$1~day is
consistent with the observed variability time scale. 
However, such a model does not immediately explain the good correlation 
between ultraviolet and X-ray variations.

\end{abstract} 


\section{ Introduction } \label{intro} 

Two of the most constraining observed properties of active galactic 
nuclei (AGN) are their large luminosities over a broad range of 
energies ($\gamma$-ray through infrared) and their rapid variability 
(implying a small source size unless the emission is highly beamed).
The inferred large energy densities have led to a standard model
of the ultimate energy source being the release of gravitational
potential energy of matter from an accretion disk surrounding
a supermassive black hole (e.g., Rees 1984).
Although this general model has broad support, the specific physical 
processes that produce the complex, broadband spectral energy distributions 
(SEDs) observed from AGN have not been clearly identified.
It is believed that a mix of processes is important.
The ultraviolet and optical emission may be primary radiation from an 
accretion disk (Shields 1978; Malkan \& Sargent 1982; Malkan 1983).
In low-luminosity objects starlight will contribute as well.
Thermal dust emission is an important ingredient of the infrared band
(Barvainis 1987; Sanders \et 1989).
The high-energy (X-ray and $\gamma$-ray)emission is not well understood.
There are a variety of models for their origin ranging from electromagnetic
cascades in an $e^+ e^-$ pair plasma (Zdziarski \et 1990) to
thermal Comptonization models (Haardt \& Maraschi 1993;
Haardt, Maraschi, \& Ghisellini 1994).
Furthermore, gas near the central source may reprocess at least some of the 
primary radiation via Compton scattering, 
absorption, and fluorescent processes (Guilbert \& Rees 1988; Lightman \& 
White 1988; George \& Fabian 1991; Matt, Fabian, \& Ross 1993).

Determination of the mix of physical processes that produce these large, 
broadband luminosities is a major unresolved issue in AGN research, and 
multi-waveband variability studies are potentially highly constraining.
Causality arguments imply that if emission in a ``secondary" band 
is produced when photons from a ````primary" band are reprocessed in 
material near the central engine, then variations in the secondary band could 
not be seen to lead those in the primary band.
Furthermore, if the emission in any given waveband is a combination of two 
independent components (with presumably independent variability behavior), 
then measurement of broadband spectral variability might allow them to be 
separated.
Finally, if a characteristic variability time scale could be measured, it 
could be compared with those indicative of different physical processes 
(e.g., with the expected viscous, orbital, light-travel time scales).

In spite of the potential power of this approach, it has not until
recently been exploited because of the very large amount of 
telescope time required.
In several experiments designed to measure the size of the broad-line region 
in the Seyfert 1 galaxy NGC 5548, variations at $\sim$1400~\AA\ were seen to 
track those at $\sim$2800~\AA\ and $\sim$5000~\AA\ to within $\ls$1--2 day 
(Clavel \et 1991; Peterson \et 1991; Korista \et 1995). 
This was taken to imply an ultraviolet-optical propagation time that is too
short to be associated with any dynamics mediated by viscosity, such as
variations in the mass inflow rate, in a standard $\alpha$-disk (Krolik \et
1991; but see \S 4.2. below). 
Similar problems were noted in ultraviolet and optical monitoring of NGC~4151
by Ulrich \et (1991).
Krolik \et (1991) suggested that variation in the different wavebands
in NGC~5548 were coordinated by a photon signal.
Nandra \et (1992) suggested that this signal might be X-ray heating
(reprocessing).  Several authors constructed specific models of
X-ray illuminated accretion disks to account for the NGC~5548 data
(Collin-Souffrin 1991; Rokaki \& Magnan 1992; Molendi, Maraschi, \& Stella
1992; Rokaki, Collin-Souffrin, \& Magnan 1993) as well as for NGC~4151
(Perola \& Piro 1994).
A strong test of the idea that the ultraviolet is produced by reprocessing 
X-ray photons could be made by measuring the time relationship between 
fluctuations in the ultraviolet and the X-rays, but previous attempts (e.g., 
Clavel \et 1992) lacked adequate temporal resolution. 

In order to attempt this test, an international consortium of AGN 
observers undertook a campaign to intensively monitor a single Seyfert~1 
galaxy, NGC~4151, at ultraviolet, X-ray, $\gamma$-ray, and optical 
wavelengths for $\sim$10 days in December 1993. 
These data are described in detail in the three preceding papers 
(Paper I--Crenshaw \et 1996; Paper II--Kaspi \et 1996; and Paper III--Warwick 
\et 1996); they are summarized in the following section.
In this paper, the multi-wavelength data are analyzed in combination.
The measurement of the multi-waveband variability, temporal correlations,
phase lags, and the broadband optical--through--$\gamma$-ray SED are 
analyzed in \S~3 and the scientific implications are briefly discussed in 
\S~4.

\section{ Data } \label{data} 

The aim of these observations was to monitor intensively NGC~4151 
across the accessible optical-through-$\gamma$-ray region during
the period from MJD 22.5 to 32.2. 
(The modified Julian Date, MJD, is defined as MJD = JD -- 2,449,300.
All dates refer to the center points of observations unless otherwise 
noted.) 
While scheduling difficulties, satellite malfunctions, and other minor
problems did cause some gaps and perturbations in this plan, this 
campaign still produced the most intensive coordinated observations
to date of an AGN across these high-energy wavebands.  
Along with daily observations before and after this period (not analyzed in 
this paper), the {\it International Ultraviolet Explorer} (\IUE) observed 
NGC~4151 nearly continuously from MJD 22.6 to 31.9.
A pair of SWP and LWP spectra, spanning the range 1200--2000~\AA\
and 2000--3000~\AA, respectively, was obtained every $\sim$70 min,
excluding a daily $\sim$2 hr gap at about modulo MJD 0.4--0.5.
\ROSAT\ observed the source twice per day, with nearly even sampling from 
MJD 22.5 to 28.0.
A satellite malfunction caused \ROSAT\ to go into safe mode for most of
the second half of the campaign, although it did make one final 
observation at MJD 32.2.
\ASCA\ was also plagued by satellite problems (this time in the first 
half of the campaign) but four 10~ksec observations were successfully 
obtained at MJD 26.0, 27.5, 29.1, and 31.6.
Compton Gamma-Ray Observatory (\CGRO) observations were made with the OSSE 
instrument during the period MJD 22.7--34.6.
Ground-based optical spectra were obtained for a 2 month period 
that included these dates.
Figure~1 gives the light curves in nominal 100~keV, 1.5~keV, 1275~\AA, 
1820~\AA, 2688~\AA\ and  5125~\AA\ wavebands.
These bands are explicitly defined below, and further details of the sampling 
characteristics are given in Table~1.

A total of 18 optical spectra were obtained during the intensive 
monitoring period; 10 at Perkins Observatory (OSU) and eight at Wise 
Observatory.
The Wise data covered 4210--6990~\AA\ with $\sim$5~\AA\ resolution through 
a 10$''$ slit, while the OSU data covered 4480--5660~\AA\ at 
$\sim$9~\AA\ resolution through a 5$''$ slit.
These data were intercalibrated to remove the effects of (presumably 
non-variable) extended emission and instrumental offsets from the light 
curves.
Continuum fluxes in the 4600$\pm$40~\AA\ and 5125$\pm$25~\AA\ bands and 
H$\beta$ line fluxes were measured from both sets of spectra, while 
6200$\pm$30~\AA\ and 6925$\pm$25~\AA\ continuum fluxes and H$\alpha$ line 
fluxes were measured only from the Wise data. 
The Wise spectrograph projects to $10'' \times 13''$ on the sky,
and approximately 25\% of the light at a wavelength of 5125~\AA\ is due
to starlight from the host galaxy (see \S~3.5.). 
The uncertainties in measurements of the optical fluxes are at about the 1\% 
level (see Paper~II for details).
The 5125~\AA\ data from both telescopes were combined in the 
correlation analyses and light curves.

During the intensive monitoring period, \IUE\ obtained a total of 
176 spectra with the SWP camera and 168 with the LWP camera.
Observations were spaced as closely as possible, leading to fairly
even sampling of $\sim$18 spectra per day in each camera, with a $\sim$2
hour period each day, during which no data were obtained 
because of Earth occultation and high particle background.
One-dimensional spectra were extracted using the TOMSIPS package
(Ayres 1993).
Continuum fluxes were measured by summing over relatively line-free, 
$\sim$30~\AA\ wide bands centered at 1275~\AA, 1330~\AA, 1440~\AA, 1820~\AA, 
1950~\AA, 2300~\AA, and 2688~\AA.
Uncertainties, taken to be the standard error in the band, were
typically 1--2\%, consistent with the observed epoch-to-epoch
dispersion in the measured fluxes.
Line fluxes (C~IV, He~II, and C~III]) were measured by fitting multiple
Gaussians.
(See Paper~I and Penton \et 1996 for details of the \IUE\ data reduction.)
The continuum light curves measured at 1275~\AA, 1820~\AA, 2688~\AA, and 
5125~\AA\ were used in the light curves and time-series analyses, all of the 
ultraviolet/optical continuum data were used only in the variability 
amplitude and zero lag correlation analyses, and the emission line data were 
not used in this paper.

The \ROSAT\ PSPC made a total of 13 observations of NGC~4151 between MJD 22.5 
and MJD 28.0. 
Although the center points of the first 12 observations were almost evenly
spaced (every 0.5~day), the integration times varied from 0.8~ksec to 
6.4~ksec.
The two soft (0.1--0.4~keV and 0.5--0.9~keV) bands, which showed no 
significant variability during the observations, are apparently dominated by 
emission from an extended component (e.g., Elvis \et 1983; Morse \et 1995).
However, the hardest \ROSAT\ band (1--2~keV) showed variations significantly 
in excess of the measured errors.

The \ROSAT\ spectra were well-fitted by a non-variable thermal bremsstrahlung 
($kT_B = 0.44 \pm 0.05 $~keV) component that contributes most of the flux 
below 1.4~keV and a heavily absorbed ($ N_H = 2.3 \times 10^{22} $~cm$^{-2}$)
power-law with fixed slope ($ \alpha = -0.5; S_\nu \propto \nu^\alpha $) and
variable normalization 
($A = 0.010 - 0.024 $ ph cm$^{-2}$ s$^{-1}$ keV$^{-1}$) that dominates at 
medium X-ray energies.
However, the proportional counter data have low resolution and thus a range 
of other models cannot be ruled out.

The four \ASCA\ spectra have higher resolution, allowing more detailed 
spectral analysis.
These data were compatible with a model that includes warm and cold
absorbers, thermal bremsstrahlung, a power-law and a 6.4~keV iron line.
Again, this is not a unique solution.
The iron line shows significant broadening to the redward wing, 
suggesting a gravitational redshift that would constrain the material
to lie very near the black hole (Yaqoob \et 1995).
The 1--2~keV \ASCA\ data were used in the light curves along with the \ROSAT\ 
points, with a 10\% uncertainty added to account for possible calibration 
differences.
The harder (2--10~keV) \ASCA\ data were not used for variability analysis 
because only four epochs were obtained.

NGC~4151 was continuously observed with the OSSE instrument on \CGRO.
The broadband 50--150~keV count rate showed weak but significant 
variability.
The raw data were calibrated by convolving the detector response function 
with a model assuming a power-law with an exponential cutoff.
See Paper~III for further details on the X-ray and $\gamma$-ray data.

NGC~4151 was near its historical peak brightness during this campaign.
The average (absorption-corrected) flux in the 2-10 keV band was 
$3.6 \times 10^{-10}~\rm~erg~cm^{-2}~s^{-1}$. 
This is bright compared to the compilation of results from {\it EXOSAT} and 
{\it Ginga} reported in Yaqoob \et (1993), which ranged from 0.8 to  
$ 4 \times 10^{-10}~\rm~erg~cm^{-2}~s^{-1}$.
The mean 1275~\AA\ flux of 
$ \sim 4 \times 10^{-13} $~erg cm$^{-2}$ s$^{-1}$ \AA$^{-1}$ is higher than 
the range of $ 0.4-3.6 \times 10^{-13} $~erg cm$^{-1}$ s$^{-1}$ \AA$^{-1}$ 
seen in 1978--1990 (Edelson, Krolik, \& Pike 1990; Courvosier \& Palatini 
1992), although the ultraviolet flux was a bit higher during the March 1995 
Astro-2 campaign (Kriss \et 1995).

\section{ Analysis } \label{analysis} 

\subsection{ Variability Amplitudes in Different Wavebands } 
\label{amplitudes} 

During the period MJD 22.5--32.3, the light curve of NGC~4151 was densely 
sampled by \IUE, \ROSAT, \CGRO, and ground-based telescopes.
The initial analysis of these data involved comparing fractional variability 
amplitudes during this period as a function of observing frequency.
The normalized variability amplitude (NVA, or $F_{var}$) was computed as 
follows.
For each band, the mean ($ \langle x \rangle $) and standard deviation 
($\sigma_{T}$) of the flux points and the mean error level ($\sigma_{err}$) 
were measured.
Because the NVA is intended to be free of instrumental effects, it was 
determined by subtracting in quadrature the measured mean error from the 
standard deviation, and then dividing by the mean flux.
That is, 
$$ F_{var}=\sqrt{{\sigma_T^2-\sigma_{err}^2}\over{\langle x \rangle^2}}. 
\eqno(1) $$
Note that this is essentially the same procedure used to derive 
$F_{var}$ in Peterson \et (1991) and $\sigma_{NVA}$ in Edelson (1992).
These quantities are given in Table~2, along with the observing band, number 
of observations ($N$) and the difference in days between the first and last 
observation ($\Delta T$) during this intensive period.

Figure~2 is a plot of NVA as a function of observing waveband.
The NVA shows a strong dependence on photon energy, increasing from 
$ F_{var} \approx 1 $\% in the optical to $\sim$4\% in the LWP and 
$ \sim 5-9 $\% in the SWP. 
The variability in the SWP is clearly stronger than at longer wavelengths, 
but because of the lower variability levels and stronger starlight 
corrections, it is impossible to say if the optical and LWP have 
significantly different ``non-stellar NVAs."
This strong wavelength dependence has been seen previously in NGC~4151 and 
other AGN (e.g., Edelson, Krolik \& Pike 1991), and is suggestive of the 
superposition of a non-variable soft component that dominates in the 
optical/infrared and a variable hard component that dominates in the 
ultraviolet.

The NVA also shows interesting behavior at high energies.
The soft \ROSAT\ bands show no evidence for any significant variations, with 
NVA formally undefined as the observed variation levels are smaller than the 
instrumental errors.
The medium energy 1.5~keV X-ray band shows the strongest variability seen 
in any band, with $ F_{var} = 24 $\%.
This large difference is apparently due to the superposition of two 
components: a soft, extended (and therefore non-variable) component seen in 
spectra and HRI images (Elvis \et 1983), and a harder, strongly variable 
component.
However, this strongly variable component must cut off at some higher 
energies, because the \CGRO\ data show weaker variability, with 
$ F_{var} \approx 6 $\% in the 50--150~keV band.

Unfortunately, there are two important energy ranges in which the variability 
properties of NGC~4151 during this campaign were not measured:
the hard X-ray 2--50~keV gap between \ROSAT\ and \CGRO\ (the 2--10~keV \ASCA\ 
data cannot be used to measure a meaningful NVA with only four points) and 
the extreme ultraviolet 1200~\AA\ (10 eV) to 100 eV gap in the mostly 
unobservable region between \IUE\ and \ROSAT.
The former waveband could contain the bulk of the luminosity of the putative 
primary emission component, and the latter, the bulk of the disk luminosity.
These gaps do limit the power of these data (see \S~4), although we note that 
this campaign has produced the most densely sampled grid in time and energy 
band obtained of any AGN to date.

Table~2 also gives the monochromatic variable luminosity, defined as 
$ L_{var} = F_{var} \times \nu L_\nu $.
($\nu L_\nu$ is the observed monochromatic luminosity, as defined in \S~3.5.)
This parameter behaves somewhat differently than the NVA.
In particular, the medium energy X-rays, which have the strongest NVA 
($F_{var} = 24$\%) actually show the lowest value of $L_{var}$
($ 6.7 \times 10^{40}$~erg s$^{-1}$).
Note however that the X-rays appear to be heavily absorbed ($\sim$90--95\%, 
see \S~3.5.), so the intrinsic value of this quantity is actually much 
larger.

\subsection{ Fluctuation Power Density Spectra } \label{PDS} 

These are the most densely sampled ultraviolet observations of any AGN 
obtained to date, allowing examination of the short time scale variability 
properties.
These data were used to measure the fluctuation power density spectra (PDS) 
in four ultraviolet bands (1275~\AA, 1330~\AA, 1820~\AA, and 2688~\AA), on 
time scales of $\sim$0.2--5~day.
A regular grid of spacing 0.1~day was created, and the value at each point 
was taken to be the average of the nearest 12 points, weighted as 
$ exp((-\Delta T/0.1day)^2) $.
This resampling, which was necessary to mitigate problems introduced by the 
periodic $\sim$0.1~day Earth occultation gaps, destroys all information on 
shorter time scales.
However, as seen below, the PDS are dominated by noise on these shorter time 
scales, so this is not a problem.
Only the continuous data were used, and the mean was not subtracted.
These resampled data were used to compute the PDS in the four ultraviolet 
bands, the results of which are shown in Figure~3.

The PDS of AGN show no discrete features that would indicate periodicity.
Instead, the fluctuation power is spread out over a wide range of temporal 
frequencies.
It is common practice to paramterize the PDS of AGN by the function 
$ P(f) \propto f^{a} $, where $P(f)$ is the PDS at temporal frequency $f$, 
and $a$ is the power-law slope.
These PDS have $ a \approx -2.5 $, making them the ``reddest" AGN PDS yet 
measured.
By comparison, the ultraviolet PDS of NGC~5548 had a slope $ a \approx -2 $  
on time scales of weeks to months (Krolik \et 1991). 
This indicates that the variability power is falling off rapidly at short 
time scales, and the bulk of the variability power is on time scales of days 
or longer.
The X-ray PDS NGC~4151 has $ a \approx -2 $ time scales of hours to days and   
$ a \approx -1 $ on longer time scales (Papadakis \& McHardy 1995).
The fact that the (rather noisy) X-ray and ultraviolet PDS appear to have 
different slopes over the same range of time scales may indicate that 
different processes power the variability at the two wavebands, although 
systematic differences in the data and reduction techniques for the two data 
sets make direct comparison difficult.

The four PDS in Figure~3 look similar to within the noise.
There do however appear to be some qualitative differences in the character 
of 
variations in different bands that can be discerned by direct examination of 
the light curves.
In particular, the ultraviolet data give the impression that the most rapid
variations occur at the shortest wavelengths, and are somewhat ``smeared out"
to longer wavelengths. 
(The $\sim$0.2~day ``spikes" in the 2688~\AA\ light curve may be due to
instrumental effects.) 
A similar effect has been seen at longer time scales in {\it HST} monitoring 
of NGC~5548 (Korista \et 1995). 
Such an effect would not be apparent in the above PDS analysis.
However, it is unclear how much of this is intrinsic and how much is due to 
instrumental differences, since the ``variability-to-noise" ratio (that is, 
$F_{var}/\sigma_{err}$), decreases toward longer wavelengths in the  
ultraviolet, so that this experiment was less sensitive to rapid variations 
at longer ultraviolet wavelengths.

\subsection{ Zero Lag Correlation Between Wavebands } \label{zerolag} 

Because there was no measurable interband phase lag (see \S~3.4), a slightly 
different view may be obtained by measuring the zero lag correlations 
between variations in different wavebands.
To investigate this, fluxes in a number of comparison bands (100~keV, 
1.5~keV, 1330~\AA, 1820~\AA, 2688~\AA, and 5125~\AA) were correlated with the 
1275~\AA\ fluxes. 
For each of the comparison bands, each flux point was paired with the average 
of the two 1275~\AA\ flux points measured immediately before and after that 
point, and 1275~\AA\ uncertainties were taken to be the average of the 
uncertainties measured for the two data points.

The results are plotted in Figure~4.
The number of points ($N$), linear correlation coefficient ($r$), Spearman 
rank correlation coefficient ($r_S$), $t$-statistic, and y-intercept are 
tabulated in Table~3.
The 100~keV fluxes ($t \approx 1$) show no correlation with 1275~\AA\ (or 
with any other band for that matter).
This could suggest that the processes producing the $\gamma$-rays are 
unrelated to those at work at lower energies, especially given the relatively 
low NVA of the $\gamma$-rays.
The data are strongly correlated on these short time scales in all other 
cases, although the optical and X-ray correlations are a bit marginal, with 
$ t = 2.95 $ for 18 points and $ t = 2.64 $ for 17 points, respectively.
Note that Perola \et (1986) found a similar ultraviolet/X-ray zero lag 
correlation in previous data over longer ($\sim$1 year) time scales, although
it was claimed that corrrelation broke down at the high flux levels seen in
this experiment.
Following the regression line to zero flux at 1275~\AA\ yields a positive
excess at all wavebands except 1.5~keV X-rays.
This means that if one would model the flux at each waveband as a combination
of a non-variable component and a variable component produced by reprocessing
emission from the other waveband, the variable component is always at shorter
wavelengths, and the non-variable component becomes a progressively larger
fraction of the total flux as the wavelength increases.

\subsection{ Multi-Wavelength Phase Lags } \label{CCFs} 

The most detailed analysis undertaken with these data was to measure temporal
cross-correlation functions between wavebands. If emission in one band is
reprocessed to another without feedback, measurement of an inter-band lag 
would provide an important confirmation, and indicate which were the primary 
and secondary bands. 
This intensive monitoring of NGC~4151 is clearly better suited for this test
than any previous campaign. 
Unfortunately, even in these data, the 1.5~keV X-rays, which show the 
strongest variations, are clearly undersampled, as they were observed at a 
temporal frequency only one-tenth that in the ultraviolet. 
The differences between the sampling, variability levels, and signal-to-noise
ratios in the different wavebands present problems for measuring the 
interband lag. 
Below, a detailed analysis is presented of two independent techniques used to 
measure the interband correlations and lags: the interpolation and discrete 
correlation functions. 

\subsubsection{ The Interpolation Cross-Correlation Method }

Cross-correlation functions were computed using the interpolation 
cross-correlation function (ICCF) method of Gaskell \& Sparke (1986),
as subsequently modified by White \& Peterson (1994). 
As described by Gaskell \& Peterson (1987), the cross-correlation
function was determined by first cross-correlating the real observations from 
one time series with values interpolated from the second series.
The calculation was then performed a second time, using the real values from
the second series and interpolated values from the first series. 
The final cross-correlation function was then taken to be the mean of these 
two calculated functions. 
A sampling grid spacing of 0.05~day was used because that is approximately 
the mean interval between the ultraviolet observations.
Because the X-ray and optical continuum light curves are much less 
well-sampled than the ultraviolet light curves, the X-ray and optical
cross-correlations with the 1275~\AA\ ultraviolet light curve were 
performed by interpolating only in the ultraviolet light curve.
That is, the computed cross-correlation functions are based on the real
X-ray and optical points and interpolated (or regularized) ultraviolet 
points, and no interpolation of the X-ray or optical data was performed. 

\subsubsection{ The Z-Transformed Discrete Correlation Function Method }

The cross-correlation functions for the various continuum bands were 
similarly calculated by the $z$-transformed discrete correlation function 
(ZDCF) method (Alexander 1996), which is related to the discrete correlation 
function (DCF) of Edelson \& Krolik (1988). 
This method is rather more general than the ICCF as (a) it does not require 
any assumptions about the continuum behavior between the actual observations, 
and (b) it is possible to reject data pairs at zero lag (e.g., two points in
different wavebands, measured from one spectrum), and thus avoid false 
zero lag correlations that might arise from  correlated flux errors.
The ZDCF differs in a number of ways from the DCF, a notable feature being 
that the data are binned by equal population rather than into time bins of 
equal width $\delta \tau$. 

One advantage of this technique is that it allows direct estimation of the 
uncertainties on the lag without the use of more assumption-dependent Monte 
Carlo simulations.
The maximum likelihood error estimate on the position of the true 
cross-correlation function peak ($\Delta\tau_{ML}$) is a 68\% fiducial 
confidence interval, meaning that it contains the peaks of 68\% of the 
likelihood-weighted population of all possible cross-correlation functions 
(Alexander 1996). 
Phrased more loosely, the confidence interval is where 68\% of the
cross-correlation functions that are consistent with the ZDCF points are 
likely
to reach their peaks. 
The accuracy of this error estimate is limited by the assumption that the
distributions of the true cross-correlation points around the ZDCF points are 
independently Gaussian, which is only approximately true, but it does not 
require any further {\it a priori} assumptions about the nature of the 
variations or transfer function between bands.

\subsubsection{ Results }

The various columns in Table 4 give the results obtained by
cross-correlating the specified light curve with the 1275~\AA\ light curve.
The parameter $r_{max}$ is the maximum value of the cross-correlation 
function, which occurs at a lag $\tau_{peak}$. 
Table~4 also gives the centroid of the cross-correlation function 
$\tau_{cent}$, which is based on all points with $r \geq 0.5r_{max}$.
The computed cross-correlation functions are shown in Figure~5 (ICCF) and 
Figure~6 (ZDCF).

The measured range of $\Delta \tau_{ML}$ of the various bands with 1275~\AA\ 
(--0.20 to +0.26 day for the 1.5~keV X-rays, --0.14 to +0.01 day 
for 1820~\AA, --0.05 to +0.22 day for 2688~\AA, --0.78 to 0.53 day for 
5125~\AA) are all consistent with zero measurable lag.
Thus, we have assigned conservative upper limits on the lags of 1275~\AA\ 
with 1.5~keV of $ \Delta \tau_{ML} \ls 0.3 $~day, with other ultraviolet 
bands of $ \Delta \tau_{ML} \ls 0.15 $~day, and with 5125~\AA\ of 
$ \Delta \tau_{ML} \ls 1 $~day.
The physical significance of these limits is discussed in \S~4.

\subsection{ Broadband Spectral Energy Distributions } \label{SEDs} 

Figure~7 gives a broadband $\gamma$-ray--optical snapshot SED, constructed by 
combining nearly-simultaneous observations consisting of the Wise spectrum 
taken on MJD 25.6, \IUE\ spectra LWP 26907 and SWP 49441, taken near MJD  
26.0, the \ROSAT\ observation centered at MJD 26.0, and 
\ASCA\ observation number 1, centered near MJD 26.0.
Because of the weak variability and low signal--to--noise in the $\gamma$-ray 
band, the \CGRO\ spectrum was integrated over the entire campaign to produce 
the high-energy SED.
Additional (non-simultaneous) infrared data from Edelson, Malkan \& Rieke 
(1987) were used in the spectral fits (\S~4.2.) but are not shown in 
Figure~7.
The data plotted are monochromatic luminosities, $\nu L_\nu$, as a function 
of frequency, $\nu$.
The quantity $\nu L_\nu$ is given by
$$ \nu L_\nu = 4 \pi D^2 \nu S_\nu , \eqno(2) $$ 
where $ S_\nu $ is the flux density and $D = 20 $~Mpc is assumed to be the 
distance to NGC~4151 (Tully 1989).

The SED of any Seyfert 1 galaxy is of course a combination of a number of 
components, and in particular, we note that starlight from the underlying Sab 
galaxy of NGC~4151 contributes a significant amount of the optical/infrared 
flux.
Peterson \et (1995) estimated that the underlying galaxy produced a flux of 
19~mJy at 5125~\AA\ in the $ 10'' \times 15'' $ Wise aperture, in good
agreement with the independent result reached in Paper~II of this series. 
This corresponds to approximately 25\% of the 5125~\AA\ flux at the light 
levels seen in this campaign, so the 5125~\AA\ NVA of the nuclear component 
(excluding starlight) is approximately 1.33 times the observed value quoted 
in Table~2.
For a normal Sab galaxy, approximately 42\% of the light at 6925~\AA, 31\% of 
the light at 6200~\AA, 21\% of the light at 4600~\AA, and 4\% of the light at 
2688~\AA\ would be due to starlight. 

Finally, we note one other important feature in the SED:
The soft X-rays show significant absorption due to line-of-sight
gas intrinsic to NGC~4151 (evident below 4~keV), and in our own Galaxy 
(evident below 0.3~keV).
Spectral fits indicate that approximately 90--95\% of the 1--2~keV luminosity 
appears to have been absorbed by gas along the line of sight (see Paper~III 
for details), although this result is model-dependent.
This would imply that the intrinsic 1--2~keV luminosity is 10--20 times that 
observed.
Possible sources of the intrinsic absorption include partial covering by cold 
gas (e.g., Holt \et 1980; Yaqoob, Warwick \& Pounds 1989; Yaqoob \et 1993) or 
absorption in a warm, ionized medium (Weaver \et 1994a,b; Warwick, Done \&
Smith 1996). 
Similar behavior has been seen in the SED of NGC~3783 obtained during the 
``World Astronomy Day" campaign (Alloin \et 1995).
A smaller fraction ($\sim$35\%) of the 2--10~keV flux is apparently absorbed, 
but unfortunately, the data are inadequate to characterize the variability in 
this band.

\subsection{ Summary of Observations} \label{summary-obs}

NGC~4151 was monitored intensively with \IUE, \ROSAT, \ASCA, \CGRO\ and 
ground-based optical telescopes for $\sim$10 days in December 1993.
These observations provided the most intensively sampled 
multi-wavelength light curve of an AGN to date.
The major new observational results are as follows:

\begin{enumerate}
\item
NGC~4151 showed strong variability from the optical through $\gamma$-ray 
bands during this period.
The optical/ultraviolet/X-ray light curves are similar but not identical, 
with no detectable lags between variations in the different bands.
The upper limits on the lags are, between 1275~\AA\ and 1.5~keV, $\ls$0.3 
day; between 1275~\AA\ and 5125~\AA, $\ls$1~day; and between 1275~\AA\ and 
the other ultraviolet bands, $\ls$0.15~day.
The 100~keV variations are not clearly related to those in any other band.
\item 
The strongest variability was seen in the medium energy X-rays, with the 
1--2~keV band showing an NVA of 24\%.
The variations were systematically weaker at lower energies, with NVAs of 
9\%, 5\%, 4\% and 0.5\% at 1275~\AA, 1820~\AA, 2688~\AA\ and 5125~\AA, 
respectively.
However, the 100~keV light curve showed an NVA of only 6\%, 
and the soft ($<$1~keV) X-rays showed no detectable variability.
The observed luminosity (as opposed to fractional) variability was seen to be 
lowest in the 1.5~keV band, apparently due to strong X-ray absorption along 
the line of sight.
\item
The PDS of all four ultraviolet bands are similar, showing no evidence for 
periodicity, but is instead being well modeled as a power-law with 
$ P(f) \propto f^{-2.5} $.
\item
The broadband SED shows a strong deficit in the soft/mid-X-ray, which is 
well-fitted as a factor of $\sim$15 absorption by gas on the line-of-sight.
\end{enumerate}

Finally, the character of the multi-wavelength variability in the
Seyfert~1 NGC~4151 appears to differ markedly from the only other AGN
monitored with similar intensity. 
The BL Lac object PKS~2155--304 was observed continually for $\sim$3 days at 
ultraviolet (Urry \et 1993), X-ray (Brinkmann \et 1994), and optical 
(Courvoisier \et 1995) wavebands.
Multi-wavelength analysis by Edelson \et (1995) found that the X-ray, 
ultraviolet and optical variations were almost identical in amplitude and 
shape, but that the X-ray variations led those in the ultraviolet and optical 
by $\sim$2--3~hr.
For these observations of NGC~4151, and other Seyfert~1s observed over longer 
time scales (e.g., Clavel \et 1991), the light curves do show significant 
differences between wavebands, with NVA a strong function of energy, but no 
lag has been measured between variations at different bands.
This is an observed (and therefore model-independent) example of an intrinsic 
difference between Seyfert~1s and BL Lacs.

\section{ Discussion } \label{discussion}

These results have important implications for models that attempt to 
explain the ultraviolet emission from AGN.
There are currently two broad classes of such models.
The first hypothesizes that the bulk of the ultraviolet luminosity is 
produced internally by viscosity in the inner regions of an accretion disk 
surrounding a central black hole, and the second, that the observed 
ultraviolet emission is produced in gas illuminated and heated by the source 
that we observe at high energies.
Of course, the true picture could be a combination of these processes,
a hybrid in which both intrinsic emission from an accretion disk and
reprocessing of X-ray emission are important (and, indeed, may feed back upon 
each other), or conversely, it may be that neither of these models is 
relevant.

\subsection{Mass and Size Scales} \label{scales}

Although the specific processes responsible for AGN emission have not been 
clearly identified, there is broad support for the general model of a black 
hole and accretion disk (see \S~1).
In this model, the emission from normal (non-blazar, radio-quiet) 
Seyfert~1s like NGC~4151 is relatively
isotropic, not significantly Doppler-boosted or beamed towards Earth.
In this case, it is possible to use variability to place relatively 
model-independent limits on the central black hole mass (and therefore the 
size scale). 

The most general is the Eddington limit, which requires only that the source 
be gravitationally bound and possess a high degree of spherical symmetry. 
The minimum central mass given by this limit is 
$$ M_E = { { L_E \sigma_e } \over { 4 \pi G c m_p } }, \eqno(3) $$
where $ L_E $ and $ M_E $ are the Eddington luminosity and mass,
$\sigma_e$ is the Thompson cross-section, and $ m_p $ is the proton mass.
As the integrated luminosity of NGC~4151 is 
$ \sim 4 \times 10^{43} $~erg s$^{-1}$,
the Eddington mass is $ M_E \approx 3 \times 10^{5} M_\odot $. 

For a source surrounding a black hole of mass $M_{BH}$, and Schwarzschild 
radius $R_S$, the minimum variability time scale ($t_{min}$) can be 
estimated from the size ($r_{min}$) of the smallest stable orbit, which is 
$$ r_{min} \approx ct_{min} \approx 3R_{S} \approx 6GM_{BH}/c^2 \eqno(4) $$
for a Schwarzschild black hole.
For $ M_{BH} \gs M_E \approx 3 \times 10^{5} M_\odot $, this implies
$ r_{min} \gs 3 \times 10^{11} $~cm or 
$ t_{min} \gs 10 $~sec, which is not a significant constraint on data 
which were sampled every hour.

A larger but more model-dependent constraint assumes Keplerian 
orbits and uses the correlation between the widths of emission lines and the 
distances estimated from their lags to estimate the central mass.
In the form in which this estimate is generally presented, the inferred
mass is the true mass if the clouds travel on circular orbits.  
If the clouds are gravitationally bound, but non-gravitational forces (e.g. 
radiation pressure or hydrodynamics) affect cloud motions, the real mass is 
greater than this estimate; if the motions are unbound, the real mass is 
smaller than this estimate.
Clavel \et (1987, 1990) used this method to derive a central mass of 
$ M \approx 4 \times 10^7 M_\odot $ for NGC~4151.
This corresponds to a smallest stable orbit of 
$ r_{min} \approx 4 \times 10^{13}~{\rm cm} \approx 20 $~lt-min.
Again, this relatively weak constraint is not significant for these data.

\subsection{Thermal Accretion Disk Models} \label{disk}

If the ultraviolet/optical continuum is optically thick thermal emission,
the variability amplitudes can be used to constrain the temperature
distribution. 
The simplest non-trivial general case is a flat, azimuthally symmetric disk 
with a local blackbody temperature that drops with radius as some power-law: 
$ T(r) = T_O (r/r_o)^{-\alpha} $.  
When $\alpha$ is 3/4, this is a fair approximation of a standard accretion 
disk, except at the smallest radii.
By contrast, in a disk that radiates predominantly by reprocessed energy (see 
\S~4.3.), $\alpha$ can be smaller, depending on the geometry of both the disk 
and the source of the primary radiation.
If the local emission is described by a blackbody, the contribution of
an annulus to the total disk flux density at a given frequency $\nu$ is
$$ S_\nu \propto \int_{x_1}^{x_2} {x^{2/\alpha-1} \over e^{-x}-1} dx, 
\eqno(5) $$
where $ x = h \nu /k T(r) $, and the starting and ending points of the 
integral are defined by the temperatures at the inner and outer radii of the 
ring.
The increasing amplitude of variations with observing frequency is then
naturally attributed to changing emission from the hotter regions (that is, 
the inner disk radii).

There is a test of the simplest case that produces simultaneous 
multi-wavelength variations, in which the emission from the outer disk 
($x > x_2$) is constant, and all of the variability is produced by a 
complete, simultaneous modulation of the emission from the inner disk 
($0 < x < x_2$).
The only free parameter is the radius that separates the variable and
constant parts of the disk, and this can be determined from the NVA at a 
single wavelength.
For example, using the 1275~\AA\ variability amplitude of 8.6\% gives
$x_2 = 0.50$ for $\alpha = 3/4$ or $ x_2 = 1.38 $ for $ \alpha = 1/2 $.
This corresponds to a boundary at the disk radius where the temperature 
is 215,000 or 78,000 K, respectively.
The last two columns of Table~2 show how the percentage flux would drop if 
all of the emission inside this radius (the putative variable component) went 
to zero, including the effects of starlight. 
There is good agreement between the simple thermal model and the observed
wavelength dependence of the variability amplitude for the standard accretion 
disk case, $ \alpha = 3/4 $. 
This is not a strong function of how the inner boundary is chosen; for the 
$ \alpha = 1/2 $ case, truncating the integration to between $ x_1 = 0.7 $ 
and $ x_2 = 1.38 $ changes the result by only 20\%.

The success of this simple exercise motivated the fitting of the
ultraviolet/optical SED with a standard model of a geometrically thin, 
optically thick accretion disk (e.g., following the formalism of Sun \& 
Malkan 1989).
Aside from the starlight, no additional long-wavelength component was 
included  in the models, so no attempt was made to fit fluxes longward 
of 2~$\mu$m.
If the disk is assumed to be viewed face-on, and the black hole is spinning
rapidly (Kerr metric), the best-fit model parameters are
$ M_{BH} = 1 \times 10^8 M_\odot $ and $ \dot M = 0.01 M_\odot$ yr$^{-1}$.
If on the other hand the black hole is assumed stationary (Schwarzschild 
case), the best-fit parameters are $ M_{BH} = 4 \times 10^7 M_\odot $ and 
$ \dot M = 0.025 M_\odot$ yr$^{-1}$.
In either case, the accretion rate corresponds to 0.6\% of the Eddington 
rate.
The black hole mass inferred from the Schwarzschild fit agrees with the 
Keplarian value obtained by Clavel \et (1987, 1990).
These fits give higher weight to the higher signal-to-noise optical continuum 
than to the ultraviolet continuum where the disk light dominates, which in 
turn requires a hotter disk and consequently a lower black hole mass 
($1 - 2 \times 10^7~\rm M_\odot$).  
All the disk fits would give significantly larger black hole masses if the 
disk had a non-zero inclination.

Integrating the multi-waveband disk emission out to a boundary radius of 
$ R \approx 2 \times 10^{14}~{\rm cm} \approx 0.07 $~lt-day shows that 
approximately 30\%, 20\% and 5\% of the total disk flux at 1275~\AA, 
2688~\AA, and 5125~\AA\ is produced in the inner disk.
(These numbers refer to the Kerr model, but are not very model-dependent;
Malkan 1991.)
After correcting for the effects of galactic starlight in the $\sim 10''$ 
aperture, the 5125~\AA\ emission from the inner disk falls to $\sim$4\%.  
For $ M_{BH} \approx 0.4 - 1 \times 10^8~M_\odot $, this radius corresponds 
to $ R \approx 6 - 15~R_S $
These fractions are approximately the same as the largest peak-to-peak 
variations observed in these bands during the intensive 10 day multi-waveband 
monitoring campaign.
Thus, one consistent explanation for the decline in variability with 
increasing wavelength is that the variations occur entirely inside the inner 
disk.
At a distance $ \approx 0.07 $ lt-day from a $ 4 \times 10^7~M_\odot $ 
black hole, the orbital time scale is $\sim$1~day, whereas the dominant 
fluctuations clearly occur on longer time scales (i.e., the 1275~\AA\
peak-to-peak variation was only 42\% over the entire 10 day intensive 
campaign).
Thus, they could be associated with either orbital mechanics or thermal 
fluctuations in the inner disk.
It must also be noted that this simple disk model does not produce any X-ray
emission, although this is also likely to originate within the inner regions.

\subsection{Reprocessing Models} \label{reprocessing}

These data also can be used to constrain models in which the ultraviolet 
radiation is produced in gas that is heated by the same X-ray continuum that 
we observe directly at higher energies. 
In this model, time variations in the ultraviolet and high energy fluxes 
should be closely coupled: the ultraviolet becomes stronger shortly after the 
high energy flux rises, and the high energy flux may itself respond to 
changes in the ultraviolet flux if the ultraviolet photons provide seeds for 
Compton upscattering into the higher energy band. 
The delays in both cases are essentially due to the light travel time
between the two source regions. 
Consequently, the (reprocessed) ultraviolet emission should vary 
simultaneously with the (primary) X-rays on time scales longer than the 
round-trip light-travel time between the emitting regions (e.g., Clavel \et 
1992). 
The strong correlation between the ultraviolet and X-ray variations 
therefore supports the reprocessing hypothesis. 
The lack of any lag within the ultraviolet could be explained in two ways:
either the sub-regions responsible for variations in different wavelengths 
are likewise very close to each other, or, as with the accretion disk model, 
only the hottest part of the region is varying. 
In the context of this model, the lack of any detectable lag implies that the
X-ray and ultraviolet emitting regions are separated by $\ls$0.15~lt-day, 
($ 1/2 \times 0.3$~lt-day, because the light must travel in both directions), 
so the bulk of the reprocessing must occur in the central regions. 
Perola \& Piro (1994) applied a more detailed reprocessing model to earlier 
X-ray/ultraviolet observations and predicted that high time resolution 
monitoring would measure lags of order 0.03--0.1~day (rather close to but 
still formally consistent with the measured limits of $\ls$0.15--0.3~day). 

The reprocessing model is supported by the broad profile of the iron 
K$\alpha$ line, which suggests relativistic effects associated with an origin 
very close to a central black hole (Yaqoob \et 1995). 
However, the \ASCA\ observations (and previous medium X-ray observations) 
found no evidence for any significant ``hard tail" in the X-ray spectrum 
(Paper III; Maisack \& Yaqoob 1991). 
While the presence of the iron line implies reprocessing by some material, 
the lack of a ``reflection hump'' suggests that the material is not optically
thick to Compton scattering, as would be expected in the putative 
reprocessing disk. 

An associated problem is the overall energy budget.
The total X-ray/$\gamma$-ray flux must be adequate to produce the 
observed ultraviolet/optical/infrared flux, and the variable 
high energy flux must also equal or exceed that at lower energies.
The observed, integrated 0.1--1~$\mu$m flux is
$ \sim 11 \times 10^{-10} $ erg cm$^{-2}$ s$^{-1}$.
As strong Ly$\alpha$ and C~IV emission indicates that the ultraviolet bump 
extends to wavelengths substantially shorter than 1000~\AA, the intrinsic,
integrated flux is probably larger by a factor of $\sim$3, 
corresponding to a total ultraviolet/optical/infrared flux of order 
$ \sim 30 \times 10^{-10} $ erg cm$^{-2}$ s$^{-1}$.
The observed, integrated 1--2~keV and 2--10~keV fluxes are 
$ \sim 0.05 \times 10^{-10} $ erg cm$^{-2}$ s$^{-1}$ and 
$ \sim 2.4  \times 10^{-10} $ erg cm$^{-2}$ s$^{-1}$, respectively
After correction for line-of-sight absorption (which is particularly 
important at 1--2~keV), the intrinsic, integrated fluxes rise to 
$ \sim 0.9 \times 10^{-10} $ erg cm$^{-2}$ s$^{-1}$ and 
$ \sim 3.6 \times 10^{-10} $ erg cm$^{-2}$ s$^{-1}$, for a total of 
$ \sim 4.5 \times 10^{-10} $ erg cm$^{-2}$ s$^{-1}$.
This is not adequate to power the lower energies, but inclusion of the GRO 
data and interpolating between ASCA and GRO yields a (rather uncertain) 
integrated 1--200~keV flux of order 
$ \sim 15 - 20 \times 10^{-10} $ erg cm$^{-2}$ s$^{-1}$.
This would be of order the amount necessary to power the lower energies.
However, the fact that the $\gamma$-ray variations differ from those in all 
other wavebands suggests that the bulk of the 10--200~keV 
emission (which dominates the X-ray/$\gamma$-ray flux) arises in a 
component that does not fully participate in the reprocessing.
In this case, the X-ray/$\gamma$-ray luminosity would still be a factor of 
order $\sim$3 too small, so at most $\sim$1/3 of the observed 
infrared/optical/ultraviolet flux could be due to reprocessing.
However, this cannot be independently verified because the \ASCA\ sampling 
above 2~keV is inadequate to characterize the variability, and indeed the 
entire 10-50~keV spectrum is interpolated, not observed.

Although the absolute X-ray luminosity changes are observed to be small 
compared to those in the ultraviolet/optical, spectral fits indicate that the 
X-rays are highly (factor of $\sim$15--20) absorbed, and the intrinsic, 
absorption-corrected variations have more than enough power to drive the 
ultraviolet/optical variations.
However, the NVA, which measures the fractional (as opposed to absolute) 
variations is 24\% in the X-rays, while it is only 9\%, 5\%, 4\%, and
1\% at 1275~\AA, 1820~\AA, 2688~\AA, and 5125~\AA, respectively.
This indicates that at most 35\%, 20\%, 15\%, and 4\% of the emission at 
1275~\AA, 1820~\AA, 2688~\AA, and 5125~\AA\ could be due to reprocessing, 
with the rest coming from a component with different (slower) variability.
As with the previous analysis, this would indicate that at most $\sim$1/3 of 
the ultraviolet/optical/infrared can be reprocessed emission from higher 
energies.

\subsection{Summary} \label{summary}

\begin{enumerate}
\item 
NGC~4151 showed significant variability over time scales of days, so the 
emitting region must be smaller than of order a few light days across,
assuming the emission is not beamed.
The limits on the interband lags indicate that, if there is reprocessing of 
flux between bands, none of the emission regions could be larger than 
$\sim$0.15~lt-day.
The lower limit to the central black hole mass derived from the Eddington 
limit is small ($ M_{BH} \gs 3 \times 10^5~M_\odot $), corresponding to a 
minimum variability time of only 10 sec, which is not a significant 
constraint.
\item 
Accretion disk fits to the SED yield a central black hole with a mass of
$ 0.4-1 \times 10^8 M_\odot $, accreting well below the Eddington limit.
The fact that the low NVAs decrease from medium energy X-ray to ultraviolet 
to optical wavebands is consistent with the accretion disk model if the bulk 
of the variable ultraviolet/optical emission originates in a region 
$\sim$0.07~lt-day ($ \sim 10 R_S $) from the center.
This model gives is no immediate explanation of the link between ultraviolet 
and X-ray variations.
\item 
The reprocessing model predicts a strong correlation between ultraviolet and 
X-ray variability, which is observed.
Because the NVAs become systematically smaller at longer wavelengths, and 
because the absorption corrected X-ray/$\gamma$-ray luminosity is only 
much smaller than in the ultraviolet/optical/infrared, and at most $\sim$1/3 
of the lower energy emission could be produced by reprocessing.
\end{enumerate}

This suggests that perhaps the ultraviolet arises in a disk powered 
partially by illumination by an X-ray source and partially by internal 
viscosity and accretion.
Determining the exact mix of these emission components will require probing 
the ultraviolet/X-ray bands at the shortest accessible time scales.
It is of particular importance to extend the coverage to the gap at X-ray 
energies harder than 2~keV.
This is the goal of the upcoming coordinated {\it XTE/IUE}/ground-based 
observations of NGC~7469, scheduled for the middle of 1996, and it is hoped 
that they will shed further light on this important question.

\acknowledgments 
The work of BMP was supported by NASA grants NAGW-3315 and NAG5-2477.
The work of DMC and the \IUE\ data reduction were supported by NASA
ADP grant S-30917-F.
The work of AVF and LCH was supported by NSF grant AST-8957063.
This research has also been supported in part by NASA grants NAG5-2439,
NAG5-1813, and NAGW-3129.

\centerline{\bf FIGURE LEGENDS} 

\noindent {\bf Figure 1} --- 
Continuum light curves during the 10 day intensive monitoring period, for 
wavebands centered on 100~keV, 1.5~keV, 1275~\AA, 1820~\AA, 2688~\AA, and 
5125~\AA.
The light curves are on a common scale, but shifted in flux to present them 
in one figure, so a 10\% flux change is shown.
The X-ray data showed the strongest variability but the poorest 
sampling; the interruption in the second half of the campaign was
due to a spacecraft malfunction.
The apparent short flares in the 2688~\AA\ light curve are 
probably due to instrumental effects, and not intrinsic variability.

\noindent {\bf Figure 2} --- 
Plot of NVA as a function of observing frequency.
Note the strong correlation in the optical/ultraviolet band,
where the NVA rises from $\sim$1\% in the optical to $\sim$9\%
in the ultraviolet.
The open circles refer to the total variability ($\sigma_T$), and the filled 
circles are the NVA ($F_{var}$), which represent the variability after 
correction for instrumental uncertainties.
The strongest NVA is seen in medium energy (1--2~keV) X-rays, with an NVA 
of 24\%, while there is no significant variability observed in the softer 
X-ray bands, and weaker variations (NVA = 6\%) at 100~keV.

\noindent {\bf Figure 3} --- 
PDS of the resampled ultraviolet light curves in four bands. 
The units of the PDS (co-ordinate axis) are erg$^2$ cm$^{-4}$ \AA$^{-2}$, 
while the units of the ordinate are day$^{-1}$.
The PDS, which all appear the same to within the errors, show no signs of 
periodic variability.
Instead, they are well-fitted by a power-law with slope $ a \approx -2.5 $.

\noindent {\bf Figure 4} --- 
Plots of the correlation between continuum fluxes at bands centered on
1275~\AA\ and 100~keV, 1.5~keV, 1330~\AA, 1820~\AA, 2668~\AA\ and
5125~\AA.  
The ultraviolet and $\gamma$-ray data are not significantly correlated, but 
all of the other bands show a significant correlation.
The solid lines are unbiased least squares fits to the data.
The regression lines with the long-wavelength ultraviolet have a positive 
y-intercept, while the ultraviolet--1.5~keV X-ray regression 
has a positive x-intercept.  

\noindent {\bf Figure 5} --- 
Interpolated cross-correlation functions between 1275~\AA\ and
1275~\AA\ (top panel; autocorrelation),
1.5~keV (second panel), 1820~\AA\ (third panel), 2688~\AA\ (fourth panel), 
and 5125~\AA\ (bottom panel).
In all cases, the peaks are consistent with zero lag.

\noindent {\bf Figure 6} --- 
Discrete cross-correlation functions between 1275~\AA\ and
1275~\AA\ (top panel; autocorrelation),
1.5~keV (second panel), 1820~\AA\ (third panel), 2688~\AA\ (fourth panel), 
and 5125~\AA\ (bottom panel).
In all cases, the peaks are consistent with zero lag.

\noindent {\bf Figure 7} --- 
Spectral energy distribution of NGC~4151.
Plotted quantities are monochromatic luminosity ($ 4 \pi D^2 \nu S_\nu $)
as a function of frequency.
Optical, ultraviolet, X-ray, and $\gamma$-ray data are taken from 
observations made near MJD 26.
The asterisks show the RMS variability in the observing bands listed in 
Table~2, in terms of monochromatic luminosity.

\clearpage 

\small
\begin{table}[h]
\begin{center}
\begin{tabular}{llccc}
\multicolumn{5}{c}{\sc Table 1} 
\\[0.2cm]
\multicolumn{5}{c}{\sc Sampling Characteristics}
\\[0.2cm]
\hline
\hline
\\
 & \multicolumn{1}{c}{$\Delta t_{ave}$}  
 & \multicolumn{1}{c}{$\Delta t_{median}$} \cr
\multicolumn{2}{l}{Light Curve}  
	& \multicolumn{1}{c}{No.\ Epochs}
	& \multicolumn{1}{c}{(days)}  
	& \multicolumn{1}{c}{(days)} 
\\[0.01cm]
\hline
& 1275~\AA, 1820~\AA\ & 176 & $0.053\pm0.022$ & 0.046 \cr
 & 2688~\AA\            & 168 & $0.055\pm0.025$ & 0.046 \cr
 & Soft X-ray            &  17 & $0.607\pm0.540$ & 0.500 \cr
 & 5125~\AA\            &  18 & $0.549\pm0.250$ & 0.585
\\[0.01cm]
\hline
\\[0.1cm]
\end{tabular}
\end{center}
\end{table}

\small
\begin{table}[h]
\begin{center}
\begin{tabular}{lcccccccc}
\multicolumn{8}{c}{\sc Table 2} 
\\[0.2cm]
\multicolumn{8}{c}{\sc Variability Amplitude vs. Band }
\\[0.2cm]
\hline
\hline
\multicolumn{1}{c}{}
 	& \multicolumn{1}{c}{Number}  
 	& \multicolumn{1}{c}{$\Delta T$}  
 	& \multicolumn{1}{c}{$\sigma_{T}$}  
 	& \multicolumn{1}{c}{$\sigma_{err}$}  
 	& \multicolumn{1}{c}{$F_{var}$}  
 	& \multicolumn{1}{c}{$L_{var}$}  
 	& \multicolumn{1}{c}{$\sigma_{3/4}$}  
 	& \multicolumn{1}{c}{$\sigma_{1/2}$}\cr
\multicolumn{1}{c}{Band}
 	& \multicolumn{1}{c}{of points}  
 	& \multicolumn{1}{c}{(days)}  
 	& \multicolumn{1}{c}{(\%)}  
 	& \multicolumn{1}{c}{(\%)} 
 	& \multicolumn{1}{c}{(\%)} 
 	& \multicolumn{1}{c}{($10^{40}$ erg s$^{-1}$)}  
 	& \multicolumn{1}{c}{(\%)}  
 	& \multicolumn{1}{c}{(\%)}
\\[0.01cm]
\hline
50-150 keV  &   9 &  8.10 &  8.9 &  6.4 &  6.1 & 119 &     &     \cr
1-2 keV     &  17 &  9.72 & 26.5 & 11.4 & 23.9 & 6.7 &     &     \cr
0.5-0.9 keV &  17 &  9.72 &  3.5 &  6.6 &  --- & --- &     &     \cr
0.1-0.4 keV &  17 &  9.72 &  3.0 &  5.0 &  --- & --- &     &     \cr
1275 \AA\   & 176 &  9.20 &  8.6 &  0.9 &  8.6 & 235 & 8.6 & 8.6 \cr
1330 \AA\   & 176 &  9.20 &  8.2 &  1.0 &  8.2 & 241 & 8.1 & 7.4 \cr
1440 \AA\   & 176 &  9.20 &  6.7 &  1.5 &  6.6 & 179 & 7.2 & 6.5 \cr
1730 \AA\   & 176 &  9.20 &  6.1 &  1.4 &  6.0 & 172 & 5.5 & 4.1 \cr
1820 \AA\   & 176 &  9.20 &  5.0 &  1.3 &  4.8 & 124 & 5.1 & 3.6 \cr
1950 \AA\   & 176 &  9.20 &  4.9 &  2.0 &  4.5 & 106 & 4.5 & 3.1 \cr
2300 \AA\   & 168 &  9.15 &  5.3 &  2.0 &  4.9 & 108 & 3.6 & 1.9 \cr
2688 \AA\   & 168 &  9.15 &  3.9 &  1.1 &  3.7 &  89 & 2.7 & 1.3 \cr
4600 \AA\   &  17 &  7.94 &  1.3 &  0.9 &  0.9 &  23 & 1.0 & 0.3 \cr
5125 \AA\   &  18 &  9.33 &  1.1 &  0.9 &  0.7 &  14 & 0.9 & 0.2 \cr
6200 \AA\   &   8 &  8.91 &  1.5 &  0.6 &  1.4 &  33 & 0.6 & 0.2 \cr
6925 \AA\   &   8 &  8.91 &  1.6 &  0.7 &  1.4 &  36 & 0.4 & 0.1
\\[0.01cm]
\hline
\\[0.1cm]
\end{tabular}
\end{center}
\end{table}
 
\clearpage

\small
\begin{table}[h]
\begin{center}
\begin{tabular}{lrccrr}
\multicolumn{6}{c}{\sc Table 3} 
\\[0.2cm]
\multicolumn{6}{c}{\sc Interband Zero Lag Correlation }
\\[0.2cm]
\hline
\hline
\multicolumn{1}{c}{Band}
 	& \multicolumn{1}{c}{N}  
 	& \multicolumn{1}{c}{$r$}  
 	& \multicolumn{1}{c}{$r_s$}  
 	& \multicolumn{1}{c}{$t$}  
 	& \multicolumn{1}{c}{y-intercept}
\\[0.01cm]
\hline
100\,keV   &   9 & 0.33 & 0.37 &  1.04 &         \cr
1\,keV     &  17 & 0.60 & 0.56 &  2.64 & $-8.76$ \cr
1330\,\AA  & 176 & 0.96 & 0.95 & 38.11 & $ 1.93$ \cr
1820\,\AA  & 176 & 0.90 & 0.86 & 22.54 & $12.11$ \cr
2688\,\AA  & 168 & 0.69 & 0.70 & 12.47 & $10.18$ \cr
5125\,\AA  &  18 & 0.65 & 0.59 &  2.95 & $ 7.21$
\\[0.01cm]
\hline
\\[0.1cm]
\end{tabular}
\end{center}
\end{table}

\small
\begin{table}[h]
\begin{center}
\begin{tabular}{lccccc}
\multicolumn{6}{c}{\sc Table 4} 
\\[0.2cm]
\multicolumn{6}{c}{\sc Cross-Correlation Results}
\\[0.2cm]
\hline
\hline
\multicolumn{1}{c}{Parameter}
 	& \multicolumn{1}{c}{Method}  
	& \multicolumn{1}{c}{Soft X-ray}
	& \multicolumn{1}{c}{UV-mid}  
	& \multicolumn{1}{c}{UV-long}
	& \multicolumn{1}{c}{Optical}\cr
 & & \multicolumn{1}{c}{(1 -- 2~kev)}  
 	& \multicolumn{1}{c}{(1820~\AA)}
	& \multicolumn{1}{c}{(2688~\AA)}
	& \multicolumn{1}{c}{(5125~\AA)}
\\[0.05cm]
\hline
$r_{max}$                &ICCF& 0.82 & 0.87 & 0.70 & 0.69 \cr
                         &ZDCF& 0.71 & 0.87 & 0.65 & 0.57 \cr
$\tau_{peak}$ (days)     &ICCF& $-0.25$ & $-0.01$ & $+0.00$ & $-0.05$ \cr
                         &ZDCF& $+0.07$ & $-0.05$ & $+0.02$ & $-0.05$ \cr
$\tau_{cent}$ (days)     &ICCF& $-0.06$ & $-0.04$ & $+0.10$ & $+0.27$ \cr
                         &ZDCF& $-0.02$ & $-0.08$ & $+0.08$ & $+0.30$ \cr
$\Delta \tau_{ML}$ (days)&ZDCF& $^{+0.19}_{-0.27}$ & $^{+0.06}_{-0.09}$ 
                              & $^{+0.20}_{-0.07}$ & $^{+0.58}_{-0.73}$ 
\\[0.01cm]
\hline
\\[0.1cm]
\end{tabular}
\end{center}
\end{table}
 
\pagestyle{empty}

\newpage

\begin{figure}
\vspace{23cm}
\includegraphics{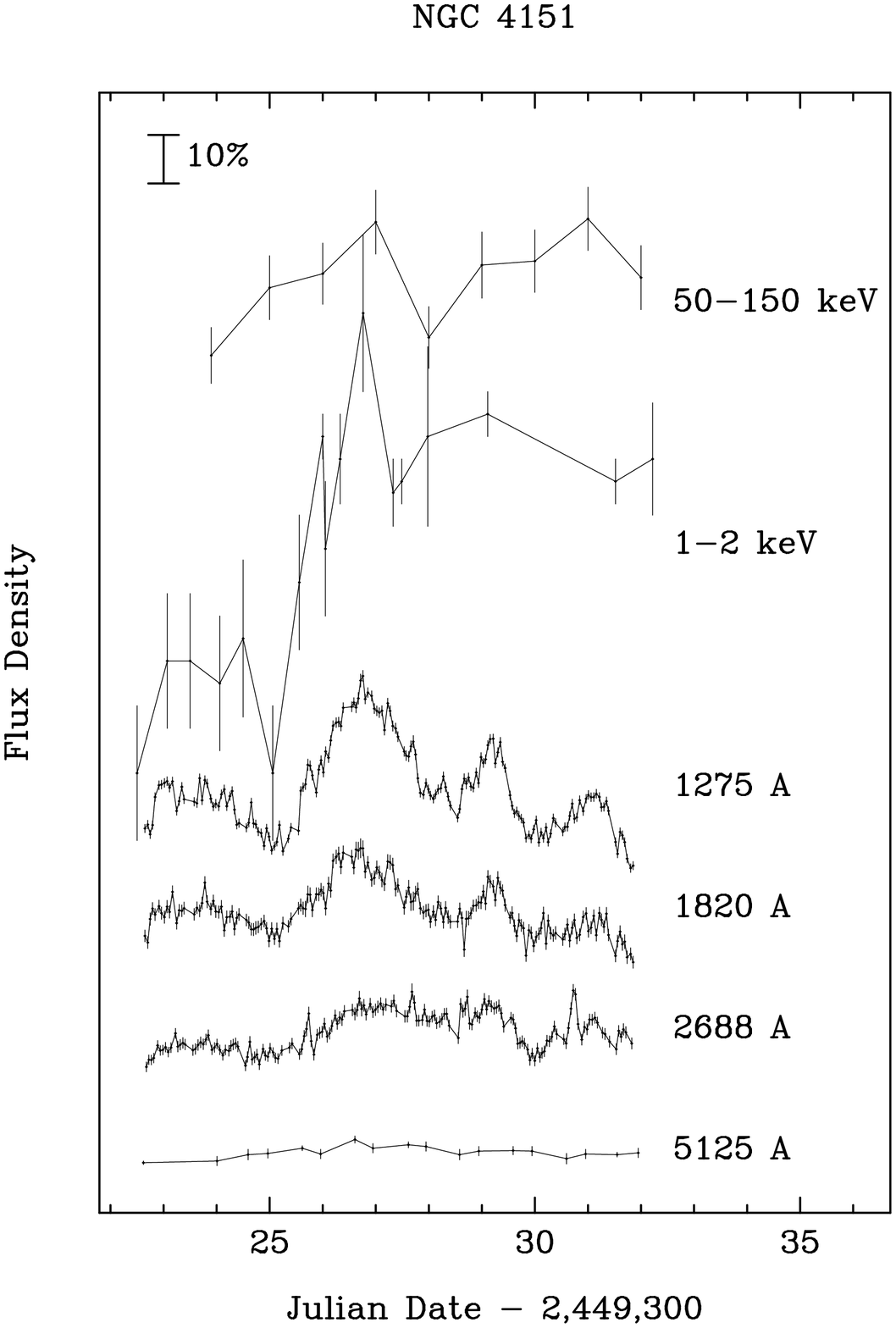}
\caption{ }
\end{figure}

\begin{figure}
\vspace{15cm}
\includegraphics{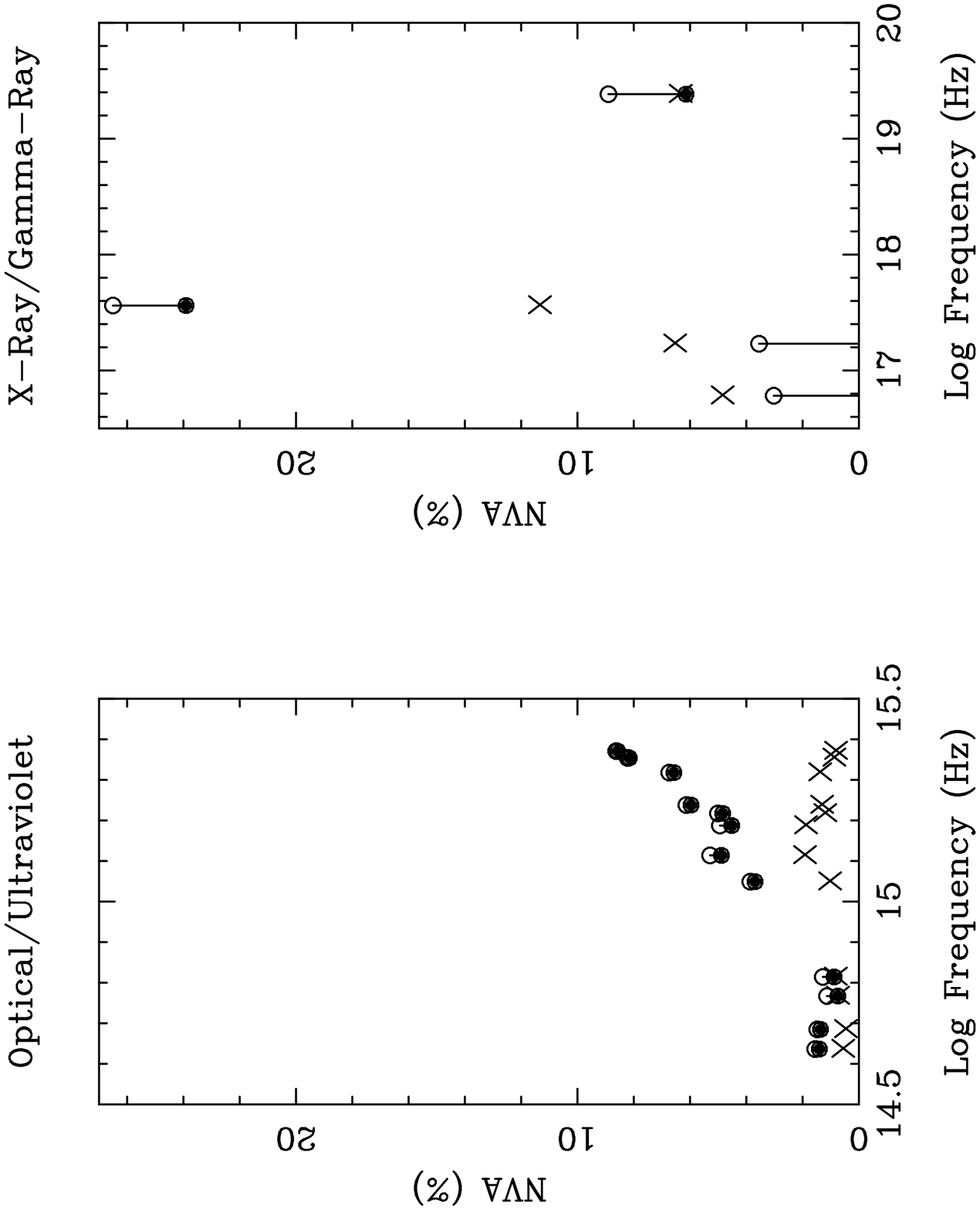}
\caption{ }
\end{figure}

\begin{figure}
\vspace{15cm}
\includegraphics{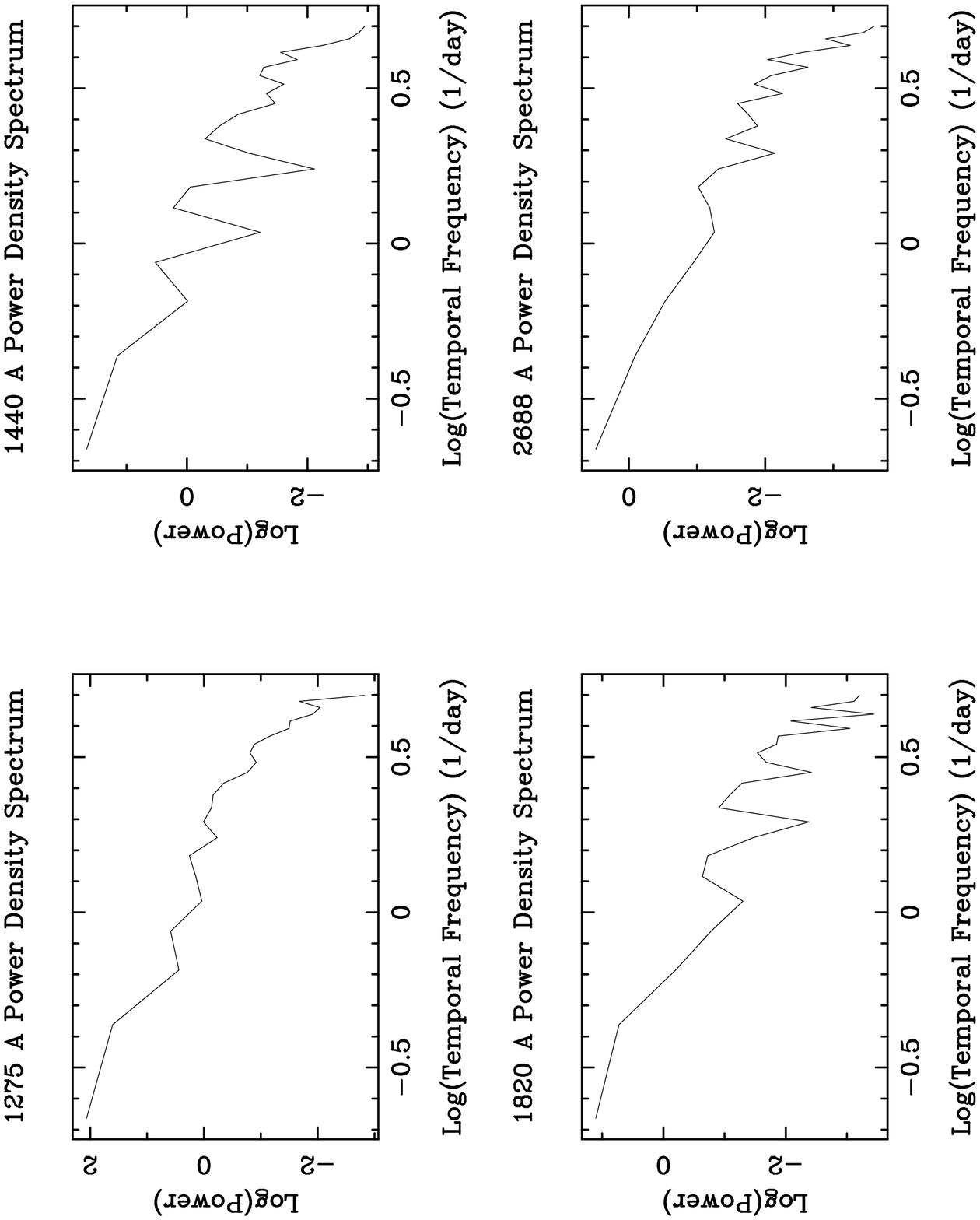}
\caption{ }
\end{figure}

\begin{figure}
\vspace{23cm}
\includegraphics{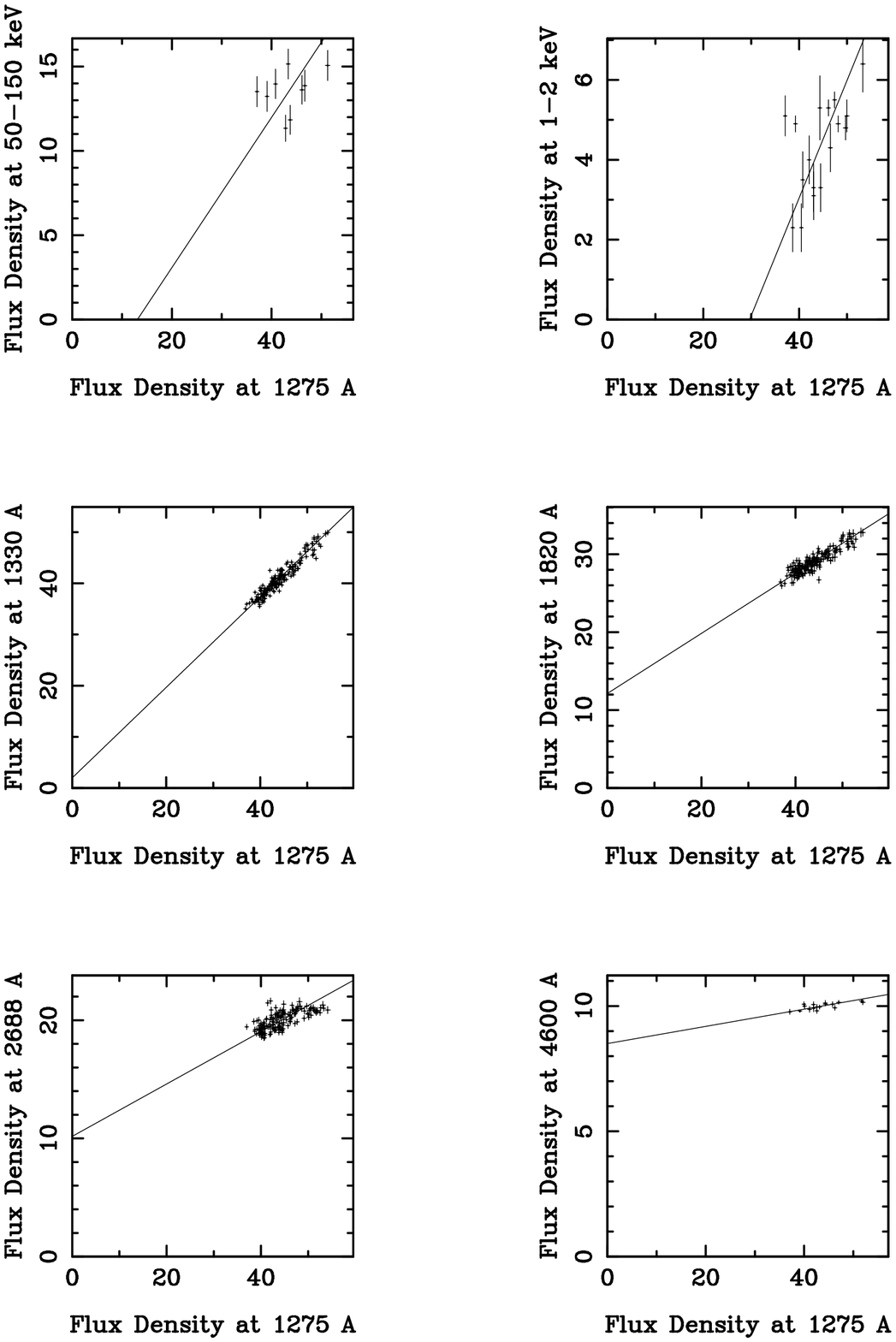}
\caption{ }
\end{figure}

\begin{figure}
\vspace{23cm}
\includegraphics{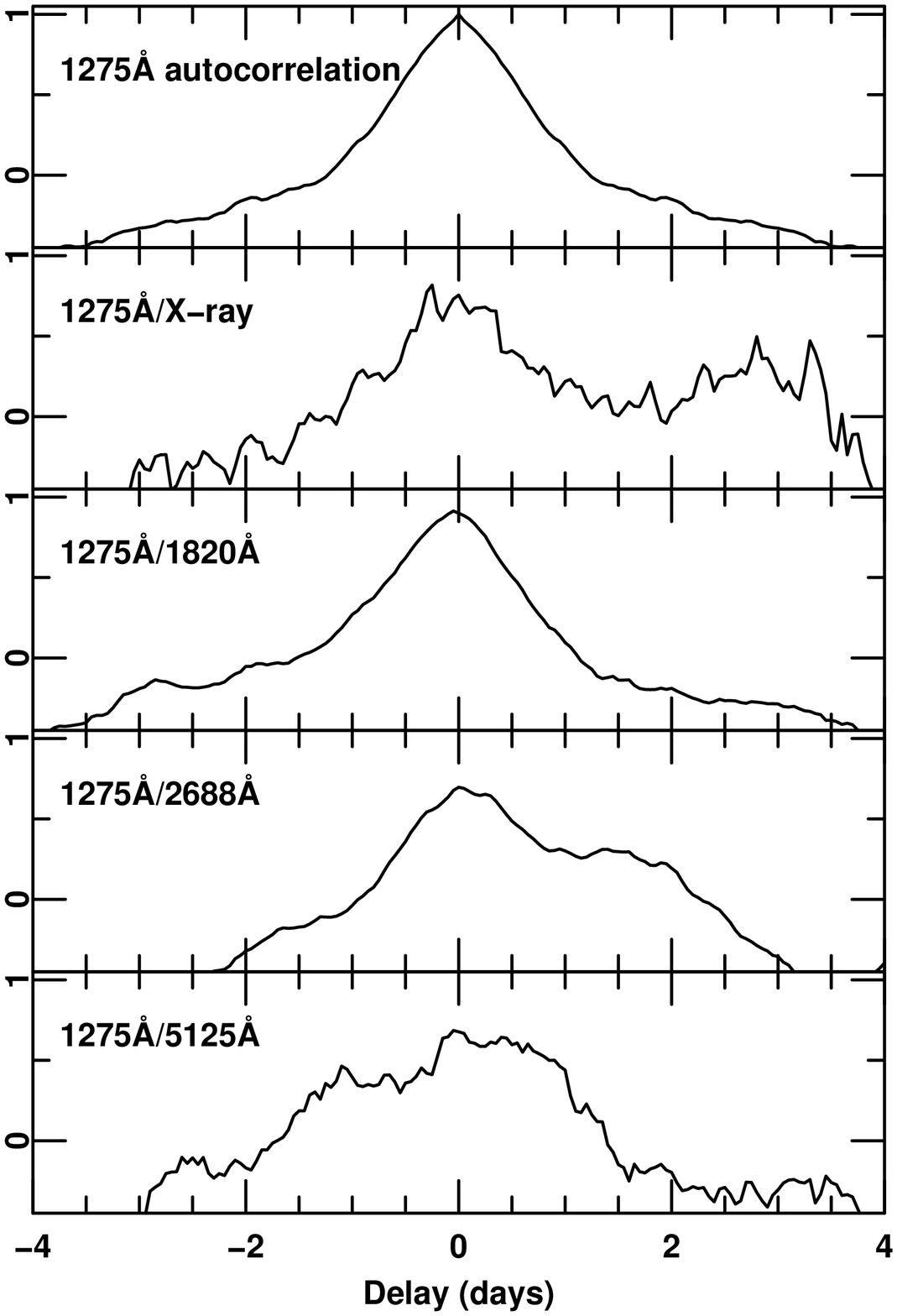}
\caption{ }
\end{figure}

\begin{figure}
\vspace{23cm}
\includegraphics{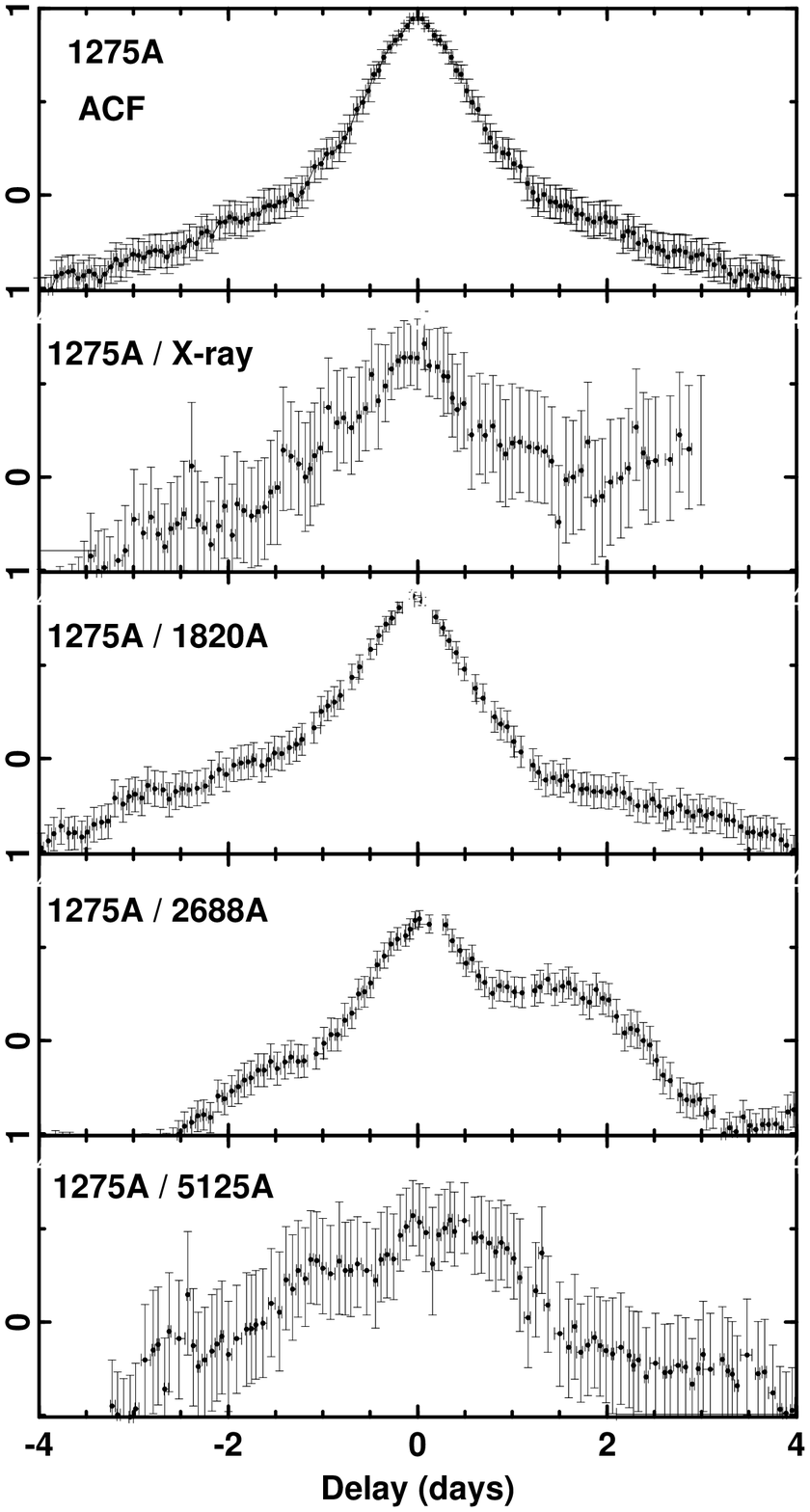}
\caption{ }
\end{figure}

\begin{figure}
\vspace{21cm}
\includegraphics{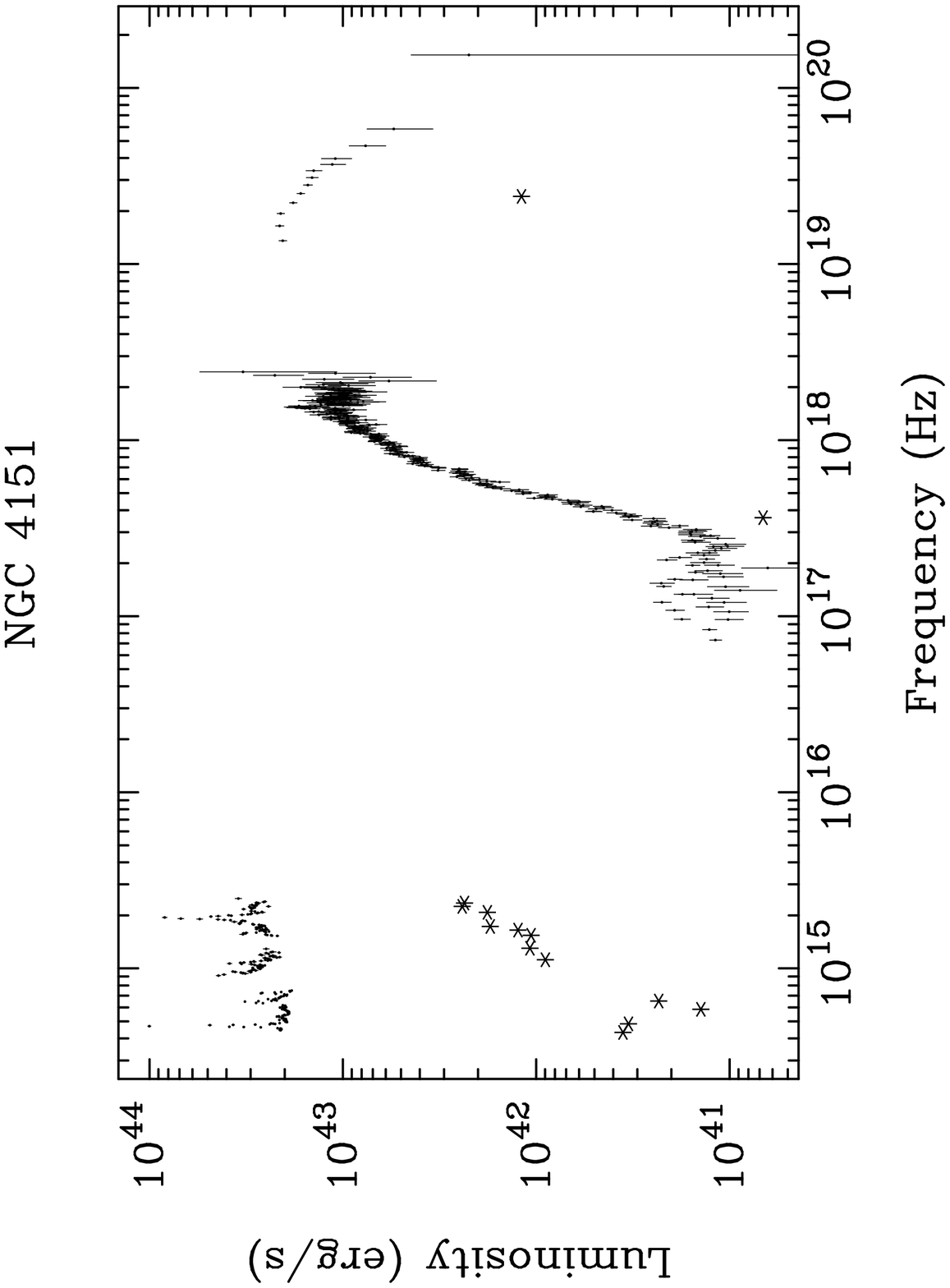}
\caption{ }
\end{figure}

\end{document}